\documentclass[Times,4pt,aps,pra,twocolumn,showpacs,amsmath,amssymb,floatfix,footinbib,superscriptaddress]{revtex4-2}
\usepackage{amsmath,natbib,graphicx,subfigure,amssymb,graphics,amsmath,mathrsfs,CJK,color,comment}
\usepackage{multirow,fancyhdr,color,bm,tabularx,psfrag,geometry,dcolumn,datetime,physics,booktabs}
\usepackage[colorlinks=true,linkcolor=blue,urlcolor=blue,citecolor=blue]{hyperref}
\usepackage[mathlines]{lineno}
\def\footnoterule{\kern -1mm \hrule width 5.8cm \kern 2.2mm}
\usepackage{tikz,xcolor,hyperref}
\definecolor{lime}{HTML}{A6CE39}
\DeclareRobustCommand{\orcidicon}{%
    \begin{tikzpicture}
    \draw[lime, fill=lime] (0,0)
    circle [radius=0.16]
    node[white] {{\fontfamily{qag}\selectfont \tiny ID}};\draw[white, fill=white] (-0.0625,0.095)
    circle [radius=0.007];
    \end{tikzpicture}
    \hspace{-2mm}}
\foreach \x in {A, ..., Z}
{\expandafter\xdef\csname orcid\x\endcsname{\noexpand\href{https://orcid.org/\csname orcidauthor\x\endcsname}{\noexpand\orcidicon}}}

\begin{document}
\title{Transient energy backflow enhanced ergotropy in an open qubit quantum battery assisted by an auxiliary oscillator}

\author{Shun-Cai Zhao\orcidA{}}
\email[Corresponding author: ]{zhaosc@kust.edu.cn.}
\affiliation{Center for Quantum Materials and Computational Condensed Matter Physics, Faculty of Science, Kunming University of Science and Technology, Kunming, 650500, PR China}

\begin{abstract}

Decoherence and dissipation in open quantum systems generally drive quantum batteries toward passive states, thereby reducing their extractable work (ergotropy). Here, we study a driven qubit quantum battery coherently coupled to an auxiliary harmonic oscillator in the presence of dephasing and dissipation. Using the differential formulation of the first law of open quantum thermodynamics, we analyze the local energy-flow dynamics during the charging process. We find that the auxiliary oscillator induces a transient negative local energy current into the qubit subsystem, corresponding to a temporary inward energy backflow. This transient energy redistribution is accompanied by an enhancement of the qubit ergotropy and a partial suppression of decoherence-induced passivation. Moreover, the parameter regime exhibiting a more pronounced negative-current interval is consistently correlated with larger ergotropy enhancement. Our results suggest that transient inward energy flow provides a useful thermodynamic signature associated with enhanced energy storage in open quantum batteries.
\begin{description}
\item[PACs]{42.50.Gy}
\item[Keywords]{quantum catalysis; qubit quantum battery; harmonic oscillator catalyst; extractable energy; external-field-driven}
\end{description}
\end{abstract}
\date{\today}
\maketitle

\section{Introduction}

The rapid development of quantum thermodynamics \cite{10.1063/1.523789,PhysRevLett.120.117702,PhysRevB.99.035421,PhysRevLett.122.210601,PhysRevE.87.042123} has stimulated extensive research on quantum batteries (QBs) \cite{PhysRevLett.132.090401,PhysRevLett.127.100601,PhysRevLett.125.236402,PhysRevB.98.205423,PhysRevLett.122.047702,d9k1-75d4} as potential nanoscale energy-storage devices \cite{PhysRevLett.131.030402,PhysRevLett.111.240401,NewJPhys.17.075015,PhysRevLett.118.150601,PhysRevA.97.022106}. The performance of a QB is commonly characterized by its \textit{ergotropy}---the maximum extractable work achievable through cyclic unitary operations \cite{PhysRevLett.129.130602,PhysRevLett.128.140501,xqtv-qbyk,PhysRevLett.133.050401}---together with its charging power. In parallel, quantum catalysis (QC) \cite{RevModPhys.96.025005,PhysRevA.107.042419,PhysRevResearch.6.023127,PhysRevLett.113.150402,PhysRevLett.85.437,PhysRevLett.83.3566}, in which an auxiliary system assists state transformations while approximately preserving its own resources \cite{PhysRevA.99.032315,PhysRevA.108.012417,PhysRevLett.122.210402,Yamasaki2022}, has attracted increasing interest in the context of QB charging dynamics.

A central challenge for practical QBs is their unavoidable coupling to dissipative environments, which gives rise to dephasing and energy relaxation \cite{PhysRevA.111.022443,qprv-gf5g,PhysRevA.111.042626,PhysRevResearch.7.L012068}. Such open-system effects tend to drive the battery toward \textit{passive states}, thereby suppressing the extractable work \cite{PhysRevB.99.035421,PhysRevLett.122.047702}. Previous studies have shown that auxiliary systems or catalytic-like mechanisms may partially mitigate this degradation through spectral restructuring, collective effects, or supermode formation \cite{PhysRevLett.132.210402,6kwv-z6fx,PhysRevResearch.6.013038,PhysRevB.108.L180301,PhysRevLett.130.020403,PhysRevE.110.044120,PhysRevA.99.032315,PhysRevA.109.052206,PhysRevA.107.042419}. However, these descriptions are often formulated at the level of global dynamics, leaving the corresponding local thermodynamic processes less transparent. In particular, the role of transient energy redistribution during the charging dynamics, and its connection to ergotropy enhancement in open systems, remains insufficiently understood.

In this work, we investigate the charging dynamics of a driven qubit QB coherently coupled to an auxiliary harmonic oscillator under simultaneous dephasing and dissipation. Our analysis focuses on the experimentally relevant energy-invariant regime, in which the auxiliary subsystem stores only a negligible amount of energy on average during the charging process \cite{PhysRevA.107.042419,RevModPhys.96.025005}. Using the differential formulation of the open-system first law ($dE=dW+dQ$), we analyze the local energy-flow dynamics of the reduced qubit subsystem.

We show that the auxiliary oscillator induces a transient negative local energy current ($J(t)<0$) into the qubit subsystem, corresponding to a temporary inward energy backflow. This transient energy redistribution is accompanied by an enhancement of the qubit ergotropy and a partial suppression of decoherence-induced passivation. Furthermore, parameter regimes exhibiting a more pronounced negative-current interval are consistently associated with larger ergotropy enhancement. These results suggest that transient inward energy flow provides a useful thermodynamic signature associated with enhanced energy storage in open quantum batteries.

The remainder of this paper is organized as follows. Sec.~II introduces the physical model and the relevant thermodynamic quantities. Sec.~III presents the charging dynamics and ergotropy enhancement induced by the auxiliary oscillator. Sec.~IV discusses the corresponding local energy-flow analysis. A possible implementation in circuit quantum electrodynamics (cQED) is outlined in Sec.~V, followed by the conclusions in Sec.~VI.

\section{Physical model}
\label{sec:model}

\begin{figure}[htbp]
\center
\includegraphics[width=0.8\columnwidth]{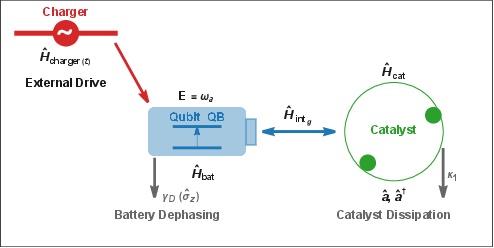}
\caption{
Schematic of the driven qubit quantum battery (QB) during the charging process. The qubit is coherently coupled to an auxiliary harmonic oscillator, while simultaneously interacting with dissipative and dephasing environments.
}
\label{fig1}
\end{figure}

We consider the charging dynamics of an open quantum battery composed of a target qubit coherently coupled to an auxiliary bosonic mode. The qubit acts as the subsystem in which useful energy is stored and extracted, whereas the auxiliary oscillator modifies the charging dynamics through coherent energy exchange. Throughout this work, the auxiliary subsystem is discussed in an operational catalytic sense, namely, as an additional degree of freedom capable of assisting the charging performance of the battery. This terminology does not imply exact catalyst-state recovery at all intermediate times, nor does it constitute a strict resource-theoretic notion of catalysis.

The total Hamiltonian is written as
\begin{align}
\hat{H}(t)
&=
\hat{H}_{0}
+
\hat{H}_{\mathrm{Cat}}
\nonumber\\
&=
\hat{H}_{\mathrm{QB}}
+
\hat{H}_{\mathrm{int}}
+
\hat{H}_{\mathrm{drive}}(t)
+
\hat{H}_{\mathrm{Cat}},
\label{H_total}
\end{align}
where $\hbar=1$. The bare Hamiltonian of the qubit battery is
\begin{equation}
\hat{H}_{\mathrm{QB}}
=
\frac{\omega_a}{2}\hat{\sigma}_z,
\label{Hqb}
\end{equation}
with transition frequency $\omega_a$. The auxiliary oscillator is modeled as a single harmonic mode,
\begin{equation}
\hat{H}_{\mathrm{Cat}}
=
\omega_c \hat{a}^{\dagger}\hat{a},
\label{Hcat}
\end{equation}
where $\omega_c$ denotes the oscillator frequency and $\hat{a}$ ($\hat{a}^{\dagger}$) is the annihilation (creation) operator. The qubit and oscillator interact through a Jaynes--Cummings coupling,
\begin{equation}
\hat{H}_{\mathrm{int}}
=
g
(
\hat{\sigma}^{+}\hat{a}
+
\hat{\sigma}^{-}\hat{a}^{\dagger}
),
\label{Hint}
\end{equation}
with coupling strength $g$. This interaction is adopted within the rotating-wave approximation and is appropriate in the near-resonant weak-to-intermediate coupling regime.

The charging process is induced by a classical drive applied to the qubit,
\begin{equation}
\hat{H}_{\mathrm{drive}}(t)
=
\Omega \sin(\omega_d t)\hat{\sigma}_x,
\label{Hdriv}
\end{equation}
where $\Omega$ and $\omega_d$ are the drive amplitude and drive frequency, respectively.

The initial state of the composite system is chosen as
\begin{equation}
\hat{\rho}(0)
=
|0\rangle_{\mathrm{QB}}\langle 0|
\otimes
|0\rangle_{\mathrm{Cat}}\langle 0|,
\label{rho0}
\end{equation}
such that the battery initially possesses zero ergotropy. The subsequent dynamics are described by the Lindblad master equation
\begin{equation}
\frac{d\hat{\rho}}{dt}
=
-i[\hat{H}(t),\hat{\rho}]
+
\mathcal{L}[\hat{\rho}],
\label{Lindmast}
\end{equation}
with dissipator
\begin{equation}
\mathcal{L}[\hat{\rho}]
=
\sum_k
\gamma_k
\left(
\hat{L}_k
\hat{\rho}
\hat{L}_k^\dagger
-
\frac{1}{2}
\{
\hat{L}_k^\dagger \hat{L}_k,
\hat{\rho}
\}
\right).
\label{Lindblad_form}
\end{equation}

We include two environmental channels: qubit dephasing and damping of the auxiliary oscillator. The corresponding jump operators are
\begin{equation}
\hat{L}_{\phi}
=
\frac{\hat{\sigma}_z}{\sqrt{2}},
\qquad
\hat{L}_{\kappa}
=
\hat{a},
\label{jump_ops1}
\end{equation}
with dephasing and damping rates $\gamma_D$ and $\kappa_1$, respectively. Under the convention of Eq.~\eqref{jump_ops1}, $\gamma_D$ directly determines the decay rate of the off-diagonal coherence of the reduced qubit density matrix.

We emphasize that local Lindblad descriptions may exhibit thermodynamic inconsistencies in asymptotic steady-state regimes under strong system-environment coupling. In the present work, however, our analysis is restricted to the transient charging regime, where the relevant dynamics occur on timescales shorter than those associated with substantial global environmental dressing. Within this regime, the local approach provides an effective description of the initial coherent energy-transfer processes relevant to the charging dynamics.

The reduced state of the qubit battery is
\begin{equation}
\hat{\rho}_{\mathrm{QB}}(t)
=
\mathrm{Tr}_{\mathrm{Cat}}
[
\hat{\rho}(t)
].
\label{rho_qb}
\end{equation}

To characterize the useful energy stored in the battery, we evaluate the ergotropy of the reduced qubit state,
\begin{equation}
\mathcal{E}(t)
=
\mathrm{Tr}
\!\left[
\hat{\rho}_{\mathrm{QB}}(t)\hat{H}_{\mathrm{QB}}
\right]
-
\mathrm{Tr}
\!\left[
\hat{\rho}_{\mathrm{passive}}(t)\hat{H}_{\mathrm{QB}}
\right],
\label{Ergotropy}
\end{equation}
where $\hat{\rho}_{\mathrm{passive}}(t)$ denotes the passive state associated with $\hat{\rho}_{\mathrm{QB}}(t)$~\cite{10.1209/epl/i2004-10101-2,PhysRevE.87.042123,PhysRevLett.127.100601}. Since our primary interest is the charging performance of the reduced qubit subsystem as a local work-storage unit, we focus on the local ergotropy rather than on global work extraction from the full composite system.

To monitor the dynamical role of the auxiliary oscillator, we additionally evaluate its instantaneous energy,
\begin{equation}
E_{\mathrm{Cat}}(t)
=
\mathrm{Tr}
\!\left[
\hat{\rho}_{\mathrm{Cat}}(t)\hat{H}_{\mathrm{Cat}}
\right],
\label{CatEnergy}
\end{equation}
where
\begin{equation}
\hat{\rho}_{\mathrm{Cat}}(t)
=
\mathrm{Tr}_{\mathrm{QB}}
[
\hat{\rho}(t)
].
\label{rho_cat}
\end{equation}

To analyze the transient charging dynamics, we further introduce the generalized local energy current associated with the qubit subsystem,
\begin{equation}
\mathcal{J}(t)
=
\mathrm{Tr}
\!\left[
\frac{d\hat{\rho}_{\mathrm{QB}}(t)}{dt}
\hat{H}_{\mathrm{QB}}
\right].
\label{enflux1}
\end{equation}

As shown in Appendix~\ref{app:energy_current_derivation}, since $\hat{H}_{\mathrm{QB}}$ is time independent, Eq.~\eqref{enflux1} can be rewritten as
\begin{equation}
\mathcal{J}(t)
=
-i
\,
\mathrm{Tr}
\!\left(
\hat{\rho}(t)
[
\hat{H}_{\mathrm{QB}},
\hat{H}_{\mathrm{int}}
+
\hat{H}_{\mathrm{drive}}(t)
]
\right)
+
\mathrm{Tr}
\!\left[
\hat{H}_{\mathrm{QB}}
\mathcal{L}[\hat{\rho}(t)]
\right].
\label{energy_current_decomp}
\end{equation}

Eq.~\eqref{energy_current_decomp} provides an operational decomposition of the local energy variation of the qubit subsystem into coherent interaction, external driving, and dissipative contributions. In the present context, $\mathcal{J}(t)$ is used as a diagnostic quantity characterizing transient local energy redistribution during the charging process, rather than as a unique thermodynamic heat current in a strict resource-theoretic sense.

All frequencies and dissipation rates are expressed in units of $\omega_a$. The master equation is solved numerically using an adaptive-step ordinary-differential-equation integrator. For the bosonic mode, a finite Fock-space truncation $N_c$ is introduced, and all reported results are verified to be converged with respect to further increases of $N_c$. Unless otherwise specified in the figure captions, the simulation parameters are summarized in Table~\ref{tab1}.

\begin{table}
\begin{center}
\vskip 0.3cm
\setlength{\tabcolsep}{5pt}
\begin{tabular}{cccccccc}
\hline
         & $\omega_{a}$ & $\omega_{c}$ & $\Omega$ & $\omega_{d}$ & $g$ & $\kappa_{1}$ & $\gamma_{D}$ \\
\hline
Fig.2,3(a) & 2.0 & / & 0.5 & 2.0 & 0.75 & 0.5 & 0.05 \\

Fig.2,3(b) & 2.0 & 0.01 & 0.5 & 2.0 & / & 0.2 & 0.05 \\

Fig.2,3(c) & 2.0 & 0.01 & 0.5 & 2.0 & 0.75 & / & 0.05 \\

Fig.2,3(d) & 2.0 & 0.01 & 0.5 & 2.0 & 0.75 & 0.2 & / \\
\hline
\end{tabular}
\caption{
Simulation parameters used in Figs.~\eqref{fig2} and~\eqref{fig3}. All frequencies and dissipation rates are expressed in consistent dimensionless units with $\hbar=1$.
}
\label{tab1}
\end{center}
\end{table}

In the following sections, we focus on how the auxiliary oscillator modifies the transient charging dynamics and the locally extractable work of the qubit battery under coherent driving and environmental dissipation.
```

\section{Results}

\subsection{Uncatalyzed ergotropy dynamics}

We first analyze the reference charging protocol in the absence of the HO catalyst. In this case, the dynamics reduce to those of a driven qubit battery subject only to pure dephasing. The corresponding open-system evolution is generated by the dephasing dissipator, which, following the structure implied by Eq.~\eqref{Lindblad_form} and standard Lindblad form, takes the form:
\begin{equation}
\mathcal{L}_{\mathrm{deph}}[\hat{\rho}]
=
\gamma_{D}
\left(
\hat{\sigma}_{z}\hat{\rho}\hat{\sigma}_{z}
-
\hat{\rho}
\right),
\label{dephasing}
\end{equation}
with jump operator $\hat{L}_{D}=\hat{\sigma}_{z}$. Here $\gamma_{D}$ denotes the qubit dephasing rate.

\begin{figure}[htbp]
\center
\includegraphics[width=0.95\columnwidth]{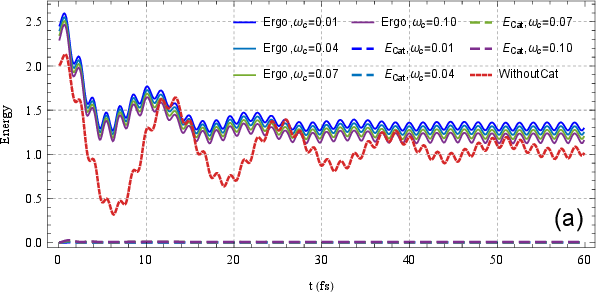 }
\includegraphics[width=0.95\columnwidth]{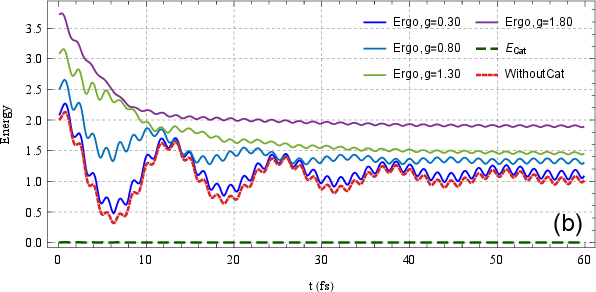 }
\includegraphics[width=0.95\columnwidth]{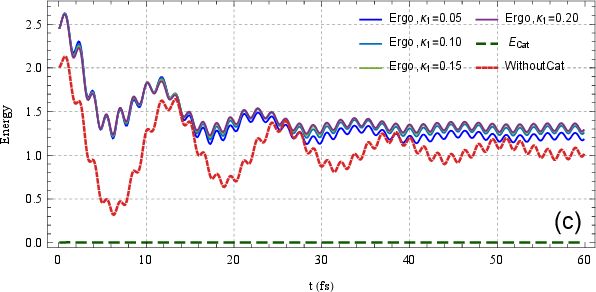 }
\includegraphics[width=0.95\columnwidth]{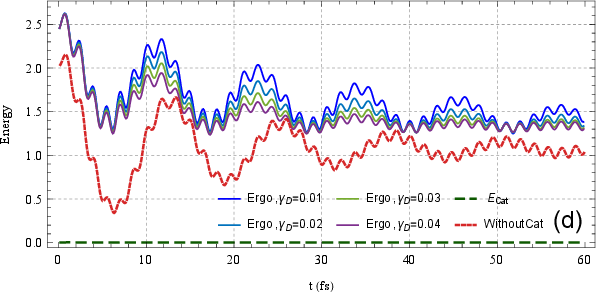 }
\caption{(Color online) Time evolution of the qubit ergotropy $\mathcal{E}(t)$ with and without the catalyst during the charging stage. Panel (a) varies the catalyst frequency $\omega_{c}$, panel (b) the qubit-catalyst coupling strength $g$, panel (c) the catalyst damping rate $\kappa_{1}$, and panel (d) the qubit dephasing rate $\gamma_{D}$. The red dashed curve in each panel denotes the uncatalyzed protocol, while the colored solid curves correspond to the catalyzed protocol. The thin dashed curve in each panel closes to zero indicate the catalyst energy $E_{\mathrm{Cat}}(t)$, which remains nearly unchanged on the scale of the figure. 
}\label{fig2}
\end{figure}

The uncatalyzed ergotropy is shown by the red dashed curves in all panels of Fig.~\ref{fig2}. Starting from the initially discharged state specified in Sec.~II, the external resonant drive first builds up extractable energy in the qubit, leading to transient oscillations in $\mathcal{E}(t)$. At later times, pure dephasing suppresses the coherence required to sustain non-passive population inversion, and the ergotropy correspondingly decreases toward a lower long-time value. Thus, even though pure dephasing does not directly remove energy from the qubit, it degrades the drive-assisted generation of non-passive states and substantially reduces the accessible extractable work.

\subsection{Catalysis-enhanced ergotropy}

We now turn to the catalyzed charging protocol. The dynamics of the coupled qubit-catalyst system are governed by the full master equation with
\begin{equation}
\mathcal{L}[\hat{\rho}]
=
\mathcal{L}_{\mathrm{cat}}[\hat{\rho}]
+
\mathcal{L}_{\mathrm{deph}}[\hat{\rho}],
\end{equation}
where
\begin{equation}
\mathcal{L}_{\mathrm{cat}}[\hat{\rho}]
=
\kappa_{1}
\left(
\hat{a}\hat{\rho}\hat{a}^{\dagger}
-
\frac{1}{2}
\left\{
\hat{a}^{\dagger}\hat{a},\hat{\rho}
\right\}
\right),
\end{equation}
and $\mathcal{L}_{\mathrm{deph}}[\hat{\rho}]$ is given by Eq.~\eqref{dephasing}.

A useful diagnostic of the catalytic regime is the oscillator energy
\begin{equation}
E_{\mathrm{Cat}}(t)
=
\mathrm{Tr}\!\left[\hat{\rho}(t)\hat{H}_{\mathrm{Cat}}\right].
\end{equation}

As indicated by the thin dashed curves in Fig.~\ref{fig2}, $E_{\mathrm{Cat}}(t)$ remains close to zero and varies only weakly throughout the evolution for the parameter range explored here. This behavior is consistent with an \emph{energy-invariant} catalytic regime \cite{PhysRevA.107.042419,RevModPhys.96.025005}, in which the auxiliary mode facilitates the charging process while storing only a negligible amount of energy on average. We stress, however, that approximate constancy of the catalyst energy does not by itself imply exact recovery of the catalyst state; the present evidence therefore supports an energetic notion of approximate catalysis rather than a strict state-invariant one.

The colored solid curves in Fig.~\ref{fig2} show that introducing the catalyst can noticeably enhance the qubit ergotropy over the uncatalyzed reference. In panels (a) and (b), the enhancement depends sensitively on both the catalyst frequency $\omega_{c}$ and the qubit-catalyst coupling strength $g$, indicating that the catalytic benefit is strongest in a restricted resonance and coupling window. Within this regime, the catalyst increases not only the peak ergotropy but also the time interval over which the battery remains appreciably non-passive.

We further examine the robustness of this enhancement against environmental noise. Figs.~\ref{fig2}(c) and \ref{fig2}(d) display the ergotropy dynamics for different values of the catalyst damping rate $\kappa_{1}$ and the qubit dephasing rate $\gamma_{D}$, respectively. Over the parameter range considered here, the catalyzed protocol consistently yields a larger ergotropy than the uncatalyzed one. This trend suggests that the auxiliary oscillator does more than simply provide an additional transient energy-exchange channel: it modifies the open-system charging dynamics in a way that helps preserve the qubit's non-passive character. To clarify this point, we next analyze the associated energy-flow decomposition.

\section{Thermodynamic analysis of catalytic enhancement}
\section{Thermodynamic analysis of catalytic enhancement}

The results presented in Sec.~III showed that the auxiliary oscillator can substantially enhance the ergotropy of the driven qubit battery over a broad parameter range. In particular, the enhancement is accompanied by a transient interval during which the qubit subsystem remains farther from passivity than in the uncatalyzed protocol. To further clarify the physical origin of this behavior, we now analyze the corresponding local energy balance within the differential formulation of open quantum thermodynamics.

Our objective here is not to claim a unique microscopic interpretation of the charging dynamics solely from subsystem thermodynamic observables. Instead, we aim to identify a consistent energetic signature associated with the catalyst-assisted preservation of non-passive qubit states during the transient charging process.

\begin{figure}[htbp]
\center
\includegraphics[width=0.95\columnwidth]{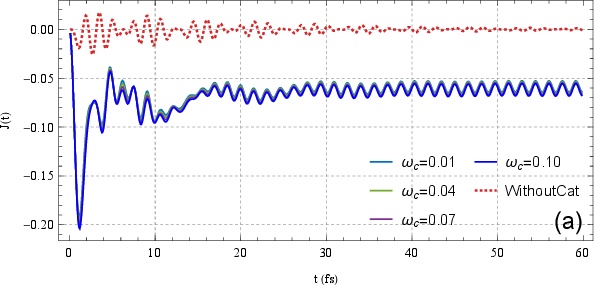 }
\includegraphics[width=0.95\columnwidth]{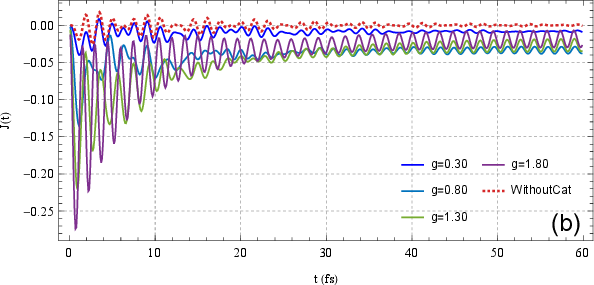 }
\includegraphics[width=0.95\columnwidth]{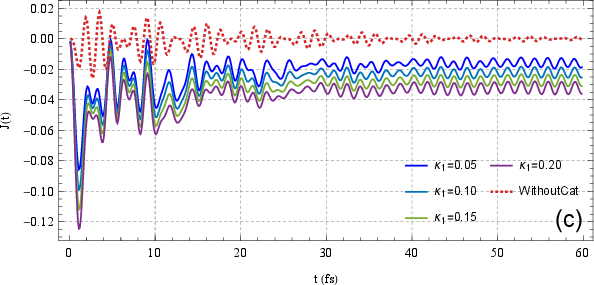 }
\includegraphics[width=0.95\columnwidth]{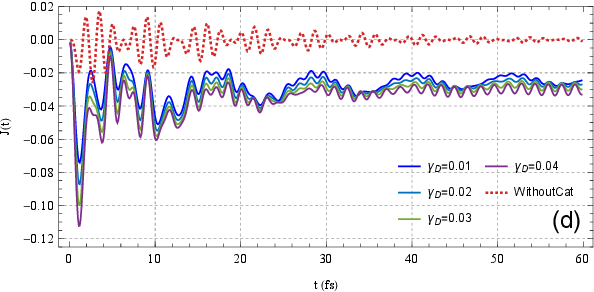 }
\caption{
(Color online) Time evolution of the local energy-current diagnostic $J(t)$ associated with the reduced qubit subsystem. The red dashed curves correspond to the uncatalyzed protocol, while the colored solid curves correspond to the catalyzed protocol. Panel (a) varies the catalyst frequency $\omega_{c}$, panel (b) the coupling strength $g$, panel (c) the catalyst damping rate $\kappa_{1}$, and panel (d) the qubit dephasing rate $\gamma_{D}$. Unless explicitly varied, the remaining parameters are fixed to the values listed in Table~\ref{tab1}. Negative values of $J(t)$ correspond to transient inward energy flow into the reduced qubit subsystem under the convention adopted in Eq.~\eqref{heatcurrent}.
}
\label{fig3}
\end{figure}

\subsection{Local energy-balance framework}

To analyze the transient charging dynamics, we define the internal energy of the reduced qubit subsystem as
\begin{equation}
E_{\mathrm{QB}}(t)
=
\mathrm{Tr}
\left[
\hat{\rho}_{\mathrm{QB}}(t)
\hat{H}_{\mathrm{QB}}
\right],
\end{equation}
where
\begin{equation}
\hat{\rho}_{\mathrm{QB}}(t)
=
\mathrm{Tr}_{\mathrm{Cat}}
[\hat{\rho}(t)]
\end{equation}
is the reduced density operator of the qubit battery.

Following the differential formulation of the first law for driven open quantum systems
\cite{1979The,1978Passive,PRXQuantum.2.030202,PhysRevE.102.062152},
the local energy variation of the reduced subsystem is written as
\begin{equation}
\frac{d}{dt}
E_{\mathrm{QB}}(t)
=
P(t)
+
J(t),
\label{FirstLaw}
\end{equation}
where
\begin{equation}
P(t)
=
\mathrm{Tr}
\left[
\hat{\rho}_{\mathrm{QB}}(t)
\frac{d\hat{H}_{\mathrm{QB}}}{dt}
\right]
\end{equation}
denotes the drive-induced power contribution, and
\begin{equation}
J(t)
=
\mathrm{Tr}
\left[
\frac{d\hat{\rho}_{\mathrm{QB}}(t)}{dt}
\hat{H}_{\mathrm{QB}}
\right]
\label{heatcurrent}
\end{equation}
defines an operational local energy-current diagnostic associated with the reduced qubit subsystem.

Since the bare qubit Hamiltonian $\hat{H}_{\mathrm{QB}}$ is time independent in the present model, the explicit driving contribution is incorporated indirectly through the reduced dynamics generated by the coherent interaction and the external drive. Using the master equation introduced in Sec.~II, Eq.~\eqref{heatcurrent} can be rewritten as
\begin{equation}
J(t)
=
-i
\,
\mathrm{Tr}
\left[
\hat{\rho}(t)
[
\hat{H}_{\mathrm{QB}},
\hat{H}_{\mathrm{int}}
+
\hat{H}_{\mathrm{drive}}(t)
]
\right]
+
\mathrm{Tr}
\left[
\hat{H}_{\mathrm{QB}}
\mathcal{L}[\hat{\rho}(t)]
\right].
\label{energy_current_analysis}
\end{equation}

Equation~\eqref{energy_current_analysis} separates the local energy variation into coherent interaction-induced and dissipative contributions under the chosen subsystem partition. In the present work, $J(t)$ is employed as a diagnostic quantity characterizing transient local energy redistribution during the charging process, rather than as a uniquely defined thermodynamic heat current in a strict microscopic sense.

With the sign convention adopted above, negative values of $J(t)$ correspond to transient inward energy flow into the reduced qubit subsystem.

We stress that such transient negative-current intervals do not by themselves imply non-Markovian reservoir memory effects. In the present setting, the effect instead originates from the coherent qubit-oscillator interaction superimposed on Markovian dissipation within the local Lindblad framework.

\subsection{Transient inward energy flow and ergotropy enhancement}

Figure~\ref{fig3} shows the time evolution of $J(t)$ for both the uncatalyzed and catalyzed charging protocols. The numerical results were obtained using the same converged integration procedure described in Sec.~II.

For the uncatalyzed protocol (red dashed curves), the reduced qubit subsystem is influenced only by the external coherent drive and pure dephasing. In this case, $J(t)$ remains close to zero throughout the evolution, with only weak oscillatory corrections. This behavior is consistent with the fact that pure dephasing does not directly induce energy relaxation for the bare qubit Hamiltonian. Its primary effect is instead the suppression of the coherence generated by the external drive, thereby reducing the generation and preservation of non-passive states.

By contrast, the catalyzed protocol exhibits a qualitatively different short-time response. In all four panels of Fig.~\ref{fig3}, the colored solid curves display a transient interval during which
\begin{equation}
J(t)<0,
\end{equation}
indicating an inward local energy flow into the reduced qubit subsystem under the adopted convention.

The magnitude and duration of this transient negative-current interval depend on the catalyst frequency $\omega_c$, the coupling strength $g$, the catalyst damping rate $\kappa_1$, and the qubit dephasing rate $\gamma_D$. Within the parameter regime explored here, parameter sets that produce a more pronounced negative-current interval also tend to generate a larger enhancement of the qubit ergotropy shown in Fig.~\ref{fig2}.

This behavior suggests that the coherent interaction with the auxiliary oscillator modifies the local energy-exchange pathways experienced by the qubit during the driven open-system evolution. As a consequence, the reduced qubit subsystem can temporarily receive energy in a manner that partially counteracts the dephasing-induced suppression of non-passivity.

At later times, the oscillation amplitude of $J(t)$ decreases and approaches values close to zero, indicating that the catalytic contribution primarily affects the transient charging stage rather than generating a continuously increasing energy accumulation process.

Comparing Figs.~\ref{fig2} and \ref{fig3}, we observe a clear dynamical correlation between the transient negative-current interval and the enhancement of the qubit ergotropy. In parameter regimes where the inward energy-flow interval becomes more pronounced, the reduced qubit subsystem generally reaches and maintains a larger ergotropy over the charging process.

We therefore interpret the transient negative-current behavior as an operational thermodynamic signature associated with the catalyst-assisted preservation of non-passive qubit states. Although this observation does not by itself provide a complete microscopic reconstruction of the charging mechanism, it offers a quantitatively accessible energetic interpretation for why the catalyzed protocol outperforms the uncatalyzed one in the present driven open-system setting.

Overall, the thermodynamic analysis supports the conclusion that the auxiliary oscillator can enhance the charging performance of the qubit battery by inducing a transient inward energy-flow interval that helps the reduced qubit subsystem remain farther from passivity in the presence of dephasing.

\section{Proposed experimental realization via circuit quantum electrodynamics}
\label{sec:experiment}

The driven open-system dynamics discussed above suggest a feasible implementation scenario within current circuit quantum electrodynamics (cQED) architectures
\cite{PhysRevA.93.013412,PhysRevB.76.144518,PhysRevA.95.022327}. In this section, we outline a possible experimental protocol for probing the transient inward energy-flow behavior associated with the enhanced ergotropy of the qubit battery.

The proposed platform consists of a superconducting transmon qubit acting as the quantum battery (QB), coherently coupled to a high-quality superconducting microwave resonator that plays the role of the auxiliary oscillator. To suppress thermal excitations, the device is assumed to operate inside a dilution refrigerator at temperatures on the order of
\begin{equation}
T \sim 10~\mathrm{mK},
\end{equation}
where the thermal occupation of the microwave mode remains small.

In a representative resonant operating regime, the characteristic frequencies may be chosen as
\begin{equation}
\omega_{\mathrm{QB}}/2\pi
\approx
\omega_{\mathrm{HO}}/2\pi
\sim
5~\mathrm{GHz}.
\end{equation}
The coherent qubit-oscillator interaction is described by the Jaynes--Cummings coupling introduced in Sec.~II. Experimentally accessible strong-coupling conditions typically correspond to
\begin{equation}
g/2\pi
\sim
50\text{--}100~\mathrm{MHz},
\end{equation}
which remain substantially larger than representative qubit dephasing rates
\begin{equation}
\gamma_D/2\pi
\sim
0.1~\mathrm{MHz}.
\end{equation}
The oscillator damping rate $\kappa_1$ may be tuned using a Purcell filter or tunable coupler, thereby allowing systematic exploration of the dissipative parameter regime relevant to the transient charging dynamics.

The experimental protocol consists of three stages: initialization, coherent charging, and state readout. First, both the qubit and the oscillator are initialized close to their ground states, preparing an approximate product state
\begin{equation}
|g\rangle_{\mathrm{QB}}
\otimes
|0\rangle_{\mathrm{HO}}.
\end{equation}
A resonant microwave pulse is then applied to the qubit for a variable charging duration $t_{\mathrm{ch}}$. Following the charging stage, the reduced density matrix of the qubit subsystem,
\begin{equation}
\hat{\rho}_{\mathrm{QB}}(t_{\mathrm{ch}}),
\end{equation}
may be reconstructed using standard time-resolved quantum state tomography (QST).

Experimentally, the tomography procedure involves repeated measurements of the Pauli observables
\begin{equation}
\langle \hat{\sigma}_x \rangle,
\qquad
\langle \hat{\sigma}_y \rangle,
\qquad
\langle \hat{\sigma}_z \rangle,
\end{equation}
typically obtained through dispersive readout techniques averaged over many repetitions. The corresponding ergotropy can then be evaluated from the reconstructed reduced density matrix using Eq.~\eqref{Ergotropy}.

In practice, the finite sampling overhead associated with QST introduces statistical uncertainty into the reconstructed density matrix, particularly at short times where the energy variation is rapid. Additional corrections may arise from measurement backaction, residual cavity leakage, finite-temperature occupation of the oscillator mode, and weak renormalization of the effective qubit frequency induced by the cavity environment. Although these effects are not expected to qualitatively alter the transient charging behavior considered here, they may influence the quantitative reconstruction of the local energy-current dynamics and should therefore be included in a detailed experimental analysis.

To probe the transient energy-flow behavior, one may estimate the operational local energy-current diagnostic introduced in Sec.~IV through the relation
\begin{equation}
J(t)
=
\frac{d}{dt}
E_{\mathrm{QB}}(t)
-
P(t),
\end{equation}
where the reduced qubit energy is obtained from
\begin{equation}
E_{\mathrm{QB}}(t)
=
\frac{\hbar \omega_{\mathrm{QB}}}{2}
\langle
\hat{\sigma}_z(t)
\rangle,
\end{equation}
and the drive-induced power contribution is determined by the externally applied microwave field.

Because numerical differentiation of experimentally reconstructed data is sensitive to shot noise and finite temporal resolution, extraction of $dE_{\mathrm{QB}}/dt$ would generally require smoothing or fitting procedures applied to the measured time series. The resulting uncertainty should therefore be interpreted together with the statistical confidence interval associated with the tomography data.

Within such an implementation, a possible experimental signature of the present mechanism would consist of the following correlated observations. First, the inferred local energy-current diagnostic exhibits a transient negative interval during the early charging stage, corresponding to inward energy flow into the reduced qubit subsystem under the adopted convention. Second, this transient interval is dynamically associated with an observable enhancement of the measured ergotropy relative to a control experiment performed without the auxiliary oscillator. Finally, partial reconstruction of the oscillator state at the end of the charging cycle may provide supporting evidence that the auxiliary mode stores only a limited amount of residual energy on average, consistent with the approximate catalytic interpretation discussed in Sec.~III.

Overall, current cQED platforms appear capable of accessing the transient parameter regime considered in this work, thereby providing a possible route for experimentally investigating the relation between transient local energy-flow dynamics and enhanced ergotropy generation in driven open quantum batteries.

\section{Conclusions}\label{sec:conclusions}

In conclusion, we have elucidated the microscopic thermodynamic mechanism driving catalysis-enhanced charging in a driven qubit quantum battery ($\text{QB}$) under open-system dynamics. By partitioning the local energy flux according to the quantum first law, we demonstrate that the catalyst functions as a coherent thermodynamic buffer. It induces a transient negative heat flux ($J(t) < 0$) that acts as a coherent energy backflow into the battery subsystem, effectively counteracting decoherence-induced passivation and driving the $\text{QB}$ into highly non-passive configurations with enhanced ergotropy.

This framework suggests that the auxiliary oscillator can induce transient inward energy redistribution within the reduced subsystem dynamics, thereby assisting the preservation of non-passive states. Future research may extend this mechanism to many-body architectures and non-Gaussian catalytic resources to explore the scaling limits of collective quantum charging and absolute thermodynamic bounds. These investigations, coupled with the proposed circuit QED implementations, will facilitate the practical deployment of high-performance quantum thermodynamic technologies.

\emph{Acknowledgments.-}
This work is supported by the National Natural Science Foundation of China ( Grant Nos. 62065009 and 61565008 ), Yunnan Fundamental Research Projects, China( Grant No. 2016FB009 ).

\section*{Code availability}\label{app:Codeavailability}

This manuscript has associated data in a data repository. [Author' comment: All data included in this manuscript are available upon reasonable request by contacting with the corresponding author]. The Supporting Information is available free of charge at: \href{https://github.com/zsczhao/data-for-Quantum-catalysis-enhanced-extract-energy-in-qubit-quantum-battery}{Codes-for-Quantum-catalysis-enhanced-extract-energy}

\appendix

\section{Derivation and interpretation of the local energy-current diagnostic}
\label{app:energy_current_derivation}

In this Appendix, we derive the operational local energy-current quantity used in the main text and clarify its physical interpretation within the reduced open-system description adopted in this work.

We emphasize from the outset that, for interacting open quantum systems, the partition of energy into subsystem, interaction, and environmental contributions is not unique. In particular, when coherent interaction terms are present, different thermodynamic decompositions may assign parts of the interaction energy differently depending on the chosen framework (e.g., local approaches, global approaches, or Hamiltonian-of-mean-force constructions). The formulation adopted here is therefore intended as an operational diagnostic of transient local energy redistribution rather than as a unique thermodynamic partition.

The reduced density matrix of the qubit battery (QB) is defined as
\begin{equation}\label{eq:app_rhoqb}
\hat{\rho}_{\mathrm{QB}}(t)=\mathrm{Tr}_{\mathrm{Cat}}\!\left[\hat{\rho}(t)\right],
\end{equation}
where $\hat{\rho}(t)$ denotes the density operator of the full qubit--oscillator system.

Following the local description introduced in the main text, we define the instantaneous internal energy of the reduced qubit subsystem with respect to the bare qubit Hamiltonian,
\begin{equation}\label{eq:app_internal_energy}
E_{\mathrm{QB}}(t)=\mathrm{Tr}\!\left[\hat{\rho}_{\mathrm{QB}}(t)\hat{H}_{\mathrm{QB}}\right],
\end{equation}
with
\begin{equation}
\hat{H}_{\mathrm{QB}}=\frac{\omega_a}{2}\hat{\sigma}_z .
\end{equation}

The corresponding local energy-current diagnostic is then introduced as
\begin{equation}\label{eq:app_Jdef}
\mathcal{J}(t)=\frac{d}{dt}E_{\mathrm{QB}}(t)=\mathrm{Tr}\!\left[\frac{d\hat{\rho}_{\mathrm{QB}}(t)}{dt}\hat{H}_{\mathrm{QB}}\right],
\end{equation}
where the equality follows from the fact that $\hat{H}_{\mathrm{QB}}$ is time independent.

The total Hamiltonian of the composite system is
\begin{align}\label{eq:app_totalH}
\hat{H}(t)&=\hat{H}_{\mathrm{QB}}+\hat{H}_{\mathrm{int}}+\hat{H}_{\mathrm{drive}}(t)+\hat{H}_{\mathrm{Cat}},
\end{align}
where
\begin{equation}
\hat{H}_{\mathrm{int}}=g\left(\hat{\sigma}^{+}\hat{a}+\hat{\sigma}^{-}\hat{a}^{\dagger}\right)
\end{equation}
describes the coherent qubit--oscillator interaction, and
\begin{equation}
\hat{H}_{\mathrm{drive}}(t)=\Omega \sin(\omega_d t)\hat{\sigma}_x
\end{equation}
is the external driving Hamiltonian.

The full dynamics obey the Lindblad master equation
\begin{equation}\label{eq:app_master}
\frac{d\hat{\rho}(t)}{dt}=-i[\hat{H}(t),\hat{\rho}(t)]+\mathcal{L}[\hat{\rho}(t)] ,
\end{equation}
where $\mathcal{L}[\cdot]$ denotes the dissipative contribution.

Taking the partial trace over the auxiliary oscillator degrees of freedom yields
\begin{equation}\label{eq:app_reduced}
\frac{d\hat{\rho}_{\mathrm{QB}}(t)}{dt}=\mathrm{Tr}_{\mathrm{Cat}}\!\left[-i[\hat{H}(t),\hat{\rho}(t)]+\mathcal{L}[\hat{\rho}(t)]\right].
\end{equation}

Substituting Eq.~\eqref{eq:app_reduced} into Eq.~\eqref{eq:app_Jdef}, and using the cyclic property of the trace together with
\begin{equation}
[
\hat{H}_{\mathrm{QB}},
\hat{H}_{\mathrm{QB}}
]=0,
\end{equation}
we obtain
\begin{align}
\mathcal{J}(t)
&=-i\,\mathrm{Tr}\!\left[\hat{\rho}(t)[\hat{H}_{\mathrm{QB}},\hat{H}_{\mathrm{int}}+\hat{H}_{\mathrm{drive}}(t)]\right]\nonumber\\
&\quad+\mathrm{Tr}\!\left[\hat{H}_{\mathrm{QB}}\mathcal{L}[\hat{\rho}(t)]\right].
\label{energy_current_decomp1}
\end{align}

Equation~\eqref{energy_current_decomp1} corresponds to Eq.~\eqref{energy_current_decomp} in the main text. The first contribution describes coherent local energy variation induced by the interaction and external driving, whereas the second contribution originates from dissipative processes contained in the Lindblad generator.

Importantly, Eq.~\eqref{energy_current_decomp1} should not be interpreted as a unique microscopic separation between work and heat in the strict thermodynamic sense. Because the qubit and oscillator are coherently coupled, the interaction energy cannot be unambiguously assigned to a single subsystem. Consequently, the quantity $\mathcal{J}(t)$ is used throughout this work as an operational local energy-current diagnostic associated with the reduced qubit dynamics under the chosen subsystem partition.

Within this interpretation, transient negative values of $\mathcal{J}(t)$ indicate intervals during which the reduced qubit subsystem experiences an effective inward local energy flow arising from the combined action of coherent interaction, external driving, and dissipation. In the main text, this quantity is employed to identify dynamical correlations between transient inward energy flow and enhanced qubit ergotropy during the charging process.

Accordingly, the quantity $\mathcal{J}(t)$ should be interpreted as an operational indicator of transient local energy redistribution under the chosen subsystem partition, rather than as a unique microscopic separation between heat and work.

\bibliography{references}
\bibliographystyle{apsrev4-2}
\end{document}